\documentstyle[eqsecnum,prb,preprint,aps,epsfig]{revtex}

\tightenlines

\newcommand{\lab}{\label}

\newcommand{\bc}{\begin{center}}
\newcommand{\ec}{\end{center}}
\newcommand{\be}{\begin{equation}}
\newcommand{\ee}{\end{equation}}
\newcommand{\bea}{\begin{eqnarray}}
\newcommand{\eea}{\end{eqnarray}}
\newcommand{\bs}{\begin{subequations}}
\newcommand{\es}{\end{subequations}}
\newcommand{\beq}{\begin{eqalignno}}
\newcommand{\eeq}{\end{eqalignno}}
\def\bol#1{\mbox{\boldmath\tiny $#1$\normalsize\unboldmath}}

\def\bol#1{{\bf #1}}

\newcommand{\half}{\frac{1}{2}}

\def\lab{\label}

\begin{document}

\draft

\title{On domain formation in non-instantaneous symmetry breaking
phase transitions}

\author{E. Alfinito$^{1,2}$ and G. Vitiello$^{1,3}$}

\address{
${}^{1}$Dipartimento di Fisica, Universit\`a di Salerno, 84100
Salerno, Italy\\
${}^{2}$ INFM, Sezione di Salerno\\
${}^{3}$INFN, Gruppo Collegato di Salerno\\
\small{alfinito@sa.infn.it,\ \ vitiello@sa.infn.it}
}

\maketitle

\begin{abstract}
In the framework of the Ginzburg-Landau harmonic
potential approximation, we present a possible modeling of 
the time-dependence of the frequency 
of the order parameter
mode suitable to account for the formation of
correlated domains in non-instantaneous symmetry breaking phase transitions.
An interesting spectrum of possibilities for the size
and the life-time of these domains emerges, which appears to 
be consistent
with the conclusions of other, differently grounded analysis.
\end{abstract}


\section{Introduction}
\label{intro}

In some recent papers \cite{BAZ,HR00,rivers}  much attention has 
been  devoted
to phase transitions in the so-called range of criticality,  namely in
the temperature range spanning between the critical temperature
$T_C$, at which the phase transition "starts", and the   Ginzburg
temperature $T_G$, at which the broken symmetry phase becomes
maximally stable. Long   string formation in such a region has been
hypothesized and its consequences have been investigated
\cite{BAZ}. Beyond the detailed  phenomenological picture and the 
numerical simulations,   the dynamic formation of
non-homogeneous structures in the critical region seems to require
further study. Our aim in this paper is to describe indeed    a
possible mechanism for the formation of domains in the process of
non-instantaneous phase transitions characterized by time dependent
order  parameter. By resorting to known results in the quantum field
theory (QFT) with spontaneous breakdown of symmetry (SBS), we also
discuss, on the basis of a microscopic analysis, the assumption of 
temperature and time dependence of the phenomenological
Ginzburg-Landau (GL) chemical potential.

To this aim let us first shortly summarize some known features of the
spontaneously broken symmetry theory with non-constant order
parameter.

We are interested in the case where a specific phase of the system at
a certain time is characterized by a specific value of  the order
parameter. Changes in the order parameter, considered as a  function
of the temperature or of the time, or of both of them, are thus
associated to a change of the system phase, i.e. to the process of
phase transition. Here   we closely follow ref.\cite{nonconst}. The
microscopic analysis may  start considering, for simplicity, a quantum
real scalar field   $\phi (x)$ at $T \neq 0$ with   Lagrangian and
Hamiltonian given by
\be \lab{1}
{\cal L} = {1 \over 2} {\partial}^{\mu} \phi \partial_{\mu} \phi -
{1 \over 2} \mu ^{2} \phi ^{2} - {1 \over 4}\lambda \phi ^{4} ~, \qquad
\lambda > 0 ~,
\ee
\be\lab{2}
H = \int d^{D-1}x [{1 \over 2} (\partial_{0} \phi)^{2} +
{1 \over 2} ({\bol \partial} \phi)^{2} + {1 \over 2} \mu ^{2} \phi ^{2}
+ {1 \over 4}\lambda \phi ^{4}] ~,
\ee
respectively, where $D$ is the space-time dimensionality and $x = ({\bf
x}, t)$. The equation of motion is
\be\lab{3}
(- \partial ^{2} - {\mu}^{2})\phi = \lambda \phi^{3} ~,
\ee
where the conventions  $g^{00}= -g^{ii}= 1$ and $c= \hbar = 1$ have
been adopted. In the variational
approach, the free energy functional is given by  the Bogoliubov
inequality \cite{Feynman}
\be\lab{4}
F \le F_{1} = F_{0} + <H-H_{0}>_{0}~,
\ee
where $F = -KT\  \ln  {\rm Tr}[\exp(-\beta H)]$ , with $\beta = {1 \over KT}$,
is the free energy and the symbol $<~>_{0}$ denotes the statistical
average: $<A>_{0} = ({1 \over Z}) {\rm Tr} [\exp(-\beta H_{0})A]$, with $Z = [\exp(-\beta H_{0})]$.
$H_{0}$ is the trial Hamiltonian. One introduces the variational parameter
${\mu_{0}}$ which in general can depend
upon $x$ and $\beta$ and acts as the mass for the quantum field $\rho$.
For a discussion on the relation between $\rho$ and the field $\phi$ see
ref. \cite{MKV86} where the detail of the variational computation is
presented. Here we only recall that we have $\phi = \rho + v$, with the
c-number $v$ being the non-vanishing expectation value
(the order parameter) for the $\phi$ field:
\be\lab{7}
<0| \phi(x)|0> = v ~.
\ee
Eq.(\ref{7}) implies $\mu^{2} < 0$ and expresses the condition of
spontaneous breakdown of symmetry. A specific physical phase of the 
system corresponds to a specific value of $v$. In the present paper we 
are interested in the case of a non-constant order parameter: $ v = v(x,
\beta) $. In such a case, Eq.(\ref{7}) tells us that the ground 
states or vacua may be
labeled by the values of the order parameter: $|0> \equiv |0>_{v}$ 
\cite{tfd,emch,sewell}. 
Thus different
vacua (i.e. different phases) correspond to different values of the 
order
parameter: $|0>_{v} \neq |0>_{v'}$, with $v' \equiv v(\beta ')$, 
$v(\beta) \neq v(\beta ')$ and $\beta \neq {\beta}'$. 
In order to study the process of phase transition one thus searches for 
the equation describing the variations of
the order parameter $v(x,
\beta)$. Such an equation may be understood as describing 
the "motion" over trajectories in the "space of the vacua" (the phases) 
of the system (also called the space of the representations of
the CCR). Let us then observe that eq. (\ref{4}) gives (we are assuming real $v(x, \beta)$)
\be
F_{1}(v(x, \beta)) = \int d^{D-1}x [(\partial_{0} v)^{2} - {\cal L}_{eff}
(v(x, \beta), \partial_{\mu} v(x,\beta))]
\lab{8}\ee
with
\be
{\cal L}_{eff} = {1 \over 2} \partial^{\mu} v \partial_{\mu} v -
U_{eff}(v) ~,
\lab{9}\ee
\be
U_{eff}(v) = U_{cl}(v) + F_{0}(\mu_{0}^{2}) - {3 \over 4}
\lambda <\rho^{2}>_{0}^{2}~,
\lab{10}\ee
\be
U_{cl}(v) = {1\over 2} \mu^{2} v^{2} + {1 \over 4} \lambda v^{4}~.
\lab{11}\ee

Note that $ U_{eff}(v)$ includes not only the classical terms but also
the quantum and the thermal contributions. We remark that the free
energy $F_{1}$ plays the r\^ole of the energy for the c-number field
$v(x, \beta)$ whose Lagrangian is ${\cal L}_{eff}$: in this way the
dynamics for the quantum field $\phi$ manifests as the dynamics for
the classical field describing the order parameter $v(x,
\beta)$. Eq. (\ref{8}) and $U_{eff}(v)$ are  the generalized
Ginzburg-Landau functional and potential, respectively. The
Euler-Lagrange equation of motion for the field $v(x, \beta)$ is
deduced from ${\cal L}_{eff}$ through the action principle:
\be
( - \partial^{2} + {m}^{2}) v(x,\beta)= \lambda
v^{3} (x,\beta) ~,
\lab{12}
\ee
with
\be\lab{13}
m^{2}  =  | \mu^{2} | - 3 \lambda <\rho^{2}>_{0}
\ee
and therefore $m^{2}$ is temperature dependent: $m^{2}=m^{2} (\beta)$.
The variational equation ${\partial F_{1} \over \partial \mu_{0}^{2}} = 0$
gives the variational mass parameter
\be \lab{14}
\mu_{0}^{2}(x, \beta) =  \mu^{2} + 3 \lambda v^{2}(x, \beta) +
3 \lambda <\rho^{2}>_{0}.
\ee
Eq.(\ref{12}) describes the evolution of the order parameter, namely
the evolution of the system in the "space of the representations of
the CCR", each representation being labeled by $v(x, \beta)$, and
therefore the evolution through the different phases of the system (phase transitions). Eq.(\ref{13}) is a consistency relation between 
the "inner"
dynamics described by   ${\cal L}$ (eq.(\ref{1})) and the dynamics in
the space of the representations described by ${\cal L}_{eff}$
(eq.(\ref{9})).  Notice that the "mass term" in eq.(\ref{12}) has
the "wrong" sign with respect to a Klein-Gordon equation for a
physical particle field: $ - m^{2}$ is interpreted as the GL chemical
potential and negative chemical potential signals the realization of
SBS. Temperature dependence in the mass term is usually introduced by
hand. Here, through eq.(\ref{13}) we see the microscopic  origin of
the temperature dependence of the chemical potential in the  GL
potential.

In the case of one space dimension one may obtain the stationary kink
solution \cite{MKV86}. When a gauge field is present and $\phi$ is a
complex field one has the vortex solution in two space dimensions. In
theories with SO(3) and SU(2) symmetry groups and three space
dimensions one has the monopole and the sphaleron solutions
\cite{MV}. A popular framework where the formation of defects in the
process of non-equilibrium symmetry breaking phase transition is
studied is the Kibble-Zurek one \cite{Kibble}.

\section{The Ginzburg regime}
\label{ginzburg}

Having presented the essential outline of the framework of our discussion,
we now turn our attention to the transition processes occurring in a finite
span of time (see, for example \cite{BAZ,HR00}) in which new phenomena, such as
the formation of exponentially decaying strings, are expected to occur.
In these processes the transition starts at the critical temperature $T_{C}$ and, after a certain lapse of time, the maximally stable configuration is attained at the so-called ``Ginzburg temperature'' \cite{Kleinbook} $T_G$~ ($T_G < T_C$). Between $T_{C}$ and $T_{G}$ the system is said to be in the critical regime \cite{BAZ}.

Our purpose is to model the time dependence
of $m^{2}(\beta)$ during the critical regime evolution, i.e.
for transitions lasting a finite time interval where the formation of
 domains is allowed.

Following ref. \cite{BAZ}, we will focus our attention on the
harmonic limit which is obtained by considering only the (linear) l.h.s.
of eq.(\ref{12}). Of course, this introduces a
strong constraint and it does not accurately describe the behavior of the system. However, it gives enough reliable information on the critical regime behavior \cite{BAZ}. We notice that in such an approximation $\mu_{0}^{2} = - m^{2}$.

Let us now consider the expansion of the $v$-field into partial waves
\be \lab{17}
v(x, \beta) = \sum_{\bol k} \ \{u_{{\bol k} }(t, \beta)
e^{i{\bol k}\cdot
{\bol x}}+u_{{\bol k}}^{\dag}(t, \beta)e^{-i{\bol k}\cdot{\bol x}
}\} ~.
\lab{18}\ee
Then, in the harmonic potential
approximation, the GL equation (\ref{12}) gives for
each $k$-mode ($k \equiv \sqrt {{\bol k}^{2}}$):
\be
\stackrel{..}{u}_{\bol k}(t)+ ({\bol k}^{2} -
m^{2})u_{\bol k}=0,\lab{19}
\ee
i.e. it leads to the equations for the parametric oscillator modes
$u_{\bol k}$ \cite{Perel} (see also \cite{AMV00}). Let us denote the
frequency of the $k$-mode by
\be
M_{k}(t)\ =\ \sqrt { {\bol k}^{2} -
m^{2}(\beta,t) }.
\lab{20}
\ee
which is required to be real for each $k$ (in this paper we are not interested in fact 
in unstable modes corresponding to imaginary 
frequency; for a recent discussion on these modes and spinodal decomposition see, e.g., ref. 
\cite{rivers} and references therein quoted). In full generality, we assume
that $m^{2}$ as well as $\beta$ may depend on time.
We remark that the reality condition on $M_{k}(t)$ turns out to be a
condition on the $k$-modes propagation. The reality condition in fact is
satisfied provided at each $t$, during the critical regime time
interval, it is
\be
{\bol k}^{2} \ge
m^{2}(\beta, t)~,
\lab{21}
\ee
for each $k$-mode. Here and in the following, for shortness we use the
notation $\beta \equiv \beta (t)$. Let $t=0$ and $t= \tau$ denote the times
at which the
critical regime starts and ends, respectively. 

For a given $k$, eq.(\ref{21}) will not hold for any $t$: of course, it
holds up to a time $t=\tau_{k}$ after which $m^{2}$ is larger than
$k^2$. Such a $\tau_k$ represents the maximal propagation time of that
$k$ mode. The value of $\tau_k$ is given when the explicit form of $m^2$
is assigned.
%
%
In the
following we consider two possibilities for the time-dependence of
$m(\beta,t)$. The first one
can lead to large correlation domains, and thus possibly excludes defect
formation. The second one does not allow very large domains and therefore
it allows defect formation. 
Our first model choice is
\be\lab{15a}
a) \hspace{3cm} \qquad m^{2}(\beta, t)  =
m_{0}^{2} \ ( e^{2f(\beta(t),t)} -1)~  , \hspace{3cm}
\ee
with $t=0$ assumed to correspond to the minimum of $m^2$ 
(we assume $f(0) = 0$ in the example of  Figure 1 and in the following).
Notice that although
at the transition temperature infinite correlation length is allowed, the
corresponding mode has only a limited time for propagating.
So the formation of domains, i.e. the 'effective causal horizon'
\cite{Kibble,Zurek}
can be inside the system (domain formation) or outside (single domain)
according to the time occurring for reaching the boundaries of the
system is longer or shorter than the allowed propagation time. So, the
dimensions to which the domains can expand depend on the rate 
between the speed at which the correlation can propagate, at a certain time, and the 
correlation length (${\lambda}_{k} \propto  (m(\beta,t))^{-1}$) at that time.

%
%
%

In the present case, each $k$-mode can
propagate for a span of time $0\le\ t \ \le \tau_k$.
From eqs.(\ref{21}) and (\ref{15a}) we obtain:
\be
f(\beta(\tau_k),\tau_k) =  \ln\left(\sqrt{k^{2} +m_{0}^{2}} \over m_{0} \right)
\quad \propto \ln\left({\cal{E}}_k\over {\cal{E}}_{0} \right)~,
\lab{22a}
\ee
where ${\cal{E}}_k$ and ${\cal{E}}_{0}$ are the $k$-mode energy for non-zero and zero $k$, respectively. 
The equilibrium time at which $T=T_G$ is $\tau\ge \tau_k$ for any $k$.

A second possibility to model $m^2$ is:
\be
b) \hspace{4cm} m^{2}(\beta, t)  =
m_{0}^{2} \  e^{2f(\beta(t),t)} ~. \hspace{4cm}
\lab{151}
\ee
In this case, a cut-off exists for the correlation length, $L \propto
m_{0}^{-1}$ (in the example of Figure 2, it corresponds to 
the time $t=0$). 
In this case the
propagation time is implicitly given by:
\be
f(\beta(\tau_k),\tau_k) =  \ln\left(k  \over m_{0} \right)
\quad \propto \ln\left( L\over \xi \right),
\lab{221}
\ee
$f(\beta(\tau_k),\tau_k)$ resembling the commonly called string tension
\cite{Zurek}. In eq. (\ref{221}) $\xi$ is the correlation length corresponding to the $k$-mode propagation. In this case the reality condition acts as an intrinsic
infrared cut-off since small $k$ values are excluded, due to
eq.(\ref{21}), by the non-zero minimum value of $m^2$. 
This means that infinitely long
wave-lengths  are actually precluded i.e. only domains of finite size
can be obtained. Finally
(phase) transitions through different vacuum states (which would be
unitarily inequivalent vacua in the infinite volume limit) at a given
$t$ are possible. This is consistent with the fact that the system is
indeed in the middle of a phase transition process (it is in the critical
regime) \cite{ARV}.  At the end of the critical regime the 
correlation may extend 
over domains of linear size of the order of $ {\tilde
{\lambda}}_{k} \propto  (m(\beta_{G},\tau))^{-1}$.

In conclusion we see that the model choice $a)$ differs from the model
choice $b)$ in the fact that the case $a)$ allows the formation of
large correlation domains (infrared modes are allowed); in the case
$b)$, on the contrary, only finite-size domains can be formed. Due of the
fact that defects may be thought as the normal symmetric phase region
trapped in between correlated (ordered) domains, we see that model $a)$
may exclude defect formation if, as mentioned above, $k$-mode life-times are
longer than the ones needed to reach the system boundaries.

%
%

%
%

We may then specify the function $f$ 
by assuming that it is positive, monotonically growing in time from $t =
0$ to $t = \tau$, at which the Ginzburg temperature is reached.


A possible analytic expression for $f(\beta(t),t)$ is
\be
f(\beta(t),t)= \
\frac{a t}{b t^{2} + c},
\lab{24} \ee
where $a, b, c$ are (positive) parameters chosen so to guarantee the
correct dimensions and the correct behavior in time and in temperature 
(see below for the relations between the various time scales and the 
ratios among $a, b, c$): $T$ decreases
from $T_ C$ to $T_G$ as time grows from $t = 0$ to $t = \tau$;
$f(\beta(t),t)$ is positive in the critical region (and even for $t >
\tau$, i.e. in the SBS region). The equilibrium time scale is given by
$\tau^{2} = {c \over b }$. Notice that in order to obtain this result,
the variations in time of $\beta(t)$ have been assumed small, according
to the picture of a slow (non-instantaneous) transition (here we refer to 
slow transitions as those for which $\tau$ is large; for fixed parameter $a$, 
large $\tau$ also means large $\tau_{Q}$, see below).
 
The behavior of $m^{2}(\beta, t) $ corresponding to (\ref{24}) is shown in
Figure 1 (case $a)$ and in Figure 2 (case $b)$.  The maximum of $m^{2} (\beta)$
has to be identified with the  minimum of the potential ${{\partial
U_{eff}} \over {\partial v}} = 0 = - m(\beta)^{2} v(x, \beta) +
\lambda v^{3} (x, \beta) $ (cf. eq.(\ref{13})), so it corresponds to
$T = T_{G}$. The negative time region in the Figures 1 and 2 
corresponds to $T > T_{C}$, so there $f$ should be negative.

Finally, the present treatment 
allows to recover 
the known results on the number of defects,
i.e. $n_{def} \propto (\tau_{0}/\tau_{Q})^{\half}$ (see
ref.\cite{rivers}, for a recent review). In fact, recalling that the equilibrium time $\tau$ is given by $\tau=\sqrt{c/b}$
it is possible to introduce the time scales: $\tau_{Q}=c/a\lambda$ and 
 $\tau_{0}=a\lambda/b$, with $\lambda$ arbitrary constant and $a,b,c$ introduced 
in eq.(\ref{24}). Thus $\tau=\sqrt{\tau_{Q}\tau_{0}}$,  which also says that for 
fixed $a$ (and $b$) large $\tau$ also means large $\tau_{Q}$. We now observe
that, at the first order approximation, one has
\be
e^{g(x)}-g(x) \approx e^{g(x_{0})}-g(x_0), \qquad {\rm if } \qquad 
\frac{\partial g(x)}{\partial x}|_{x_0}=0,
\lab{exp1}\ee
so, in our case, $e^{2f(\tau)}=2f(\tau)+e^{2f(0)}$.


The number of defects can be finally obtained. Our choice is to
consider the defect number at $T_{G}$, namely when the system gets
enough stabilized (in some sense we count the defects that are still
present at the end of the critical regime). In the case $a)$ it is: %
\be n_{def} \propto m^{2}(\beta_{G},\tau)\approx 2m_{0}^{2} f(\tau)
 =m_{0}^{2}\ \tau/{\lambda \tau_{Q}} \propto
\sqrt{(\tau_{0}/\tau_{Q})}.  
\lab{24bis} \ee 
Similarly, in the case
$b)$: 
\be n_{def} \propto m^{2}(\beta_{G},\tau)\approx m_{0}^{2}
(2f(\tau)+1) \approx m_{0}^{2}\ \tau/{\lambda \tau_{Q}}\qquad 
{\rm for\; large \; \tau} ~~ {\rm (slow\; transitions) } 
\lab{24ter} \ee 
As observed above,
for fixed $a$ (and $b$) large $\tau$ also means large $\tau_{Q}$. We
note that the form of $m^{2}$ is a good approximation at the Ginzburg
temperature in our models, since, as already observed, at $T_{G}$ it
corresponds to the minimum of the potential, which is where we want to
know the defect number. It is in fact interesting that Zurek's results
\cite{Zurek} are there also recovered.

\section{Final remarks}

It is interesting to write $M_{k}^{2}(t)$ as 
\be
M_{k}^{2}(\Lambda_{k} (t))\ =\ M_{k}^{2}(0)\ \ e^{-2 \Lambda_{k}(t)},
\lab{25} \ee
%
\be e^{- 2 \Lambda_{k} (t)} =
e^{2f(\beta(t),t)} \ \frac{\sinh(f(\tau_{k})-f(\beta(t),t))}{\sinh
f(\tau_{k})}, \lab{26}
\ee
where $f(0) = 0$ has been used. Eq.(\ref{26}) shows that
$\Lambda_{k} (t)\ge 0$ for $0 \le t \le \tau_{k}$ , $\Lambda_{k}
(0)=0$ and $\Lambda_{k}(\tau_{k})=\infty$.  Since $ M_{k}(\Lambda_{k}
(\tau_{k}))\ = 0$ we see that $\Lambda_{k} (t)$ acts as a life-time,
say with $\Lambda_{k} (t) \propto s_{k}$, for the   $k$-mode: each
$k$-mode "lives" with a proper time $s_{k}$, i.e. it is born when
$s_{k}$ is zero and it dies for $s_{k} \rightarrow \infty$. 
The "lives" of the
$k$-modes for growing $k$ are shown in Figure 3, where
$\Lambda_{k} (t)$ is plotted versus $t$ for different values of
$k$. The blowing up values are reached in correspondence of the
$\tau_{k}$ values for each $k$-mode.
Only the
modes satisfying the reality condition are present at a certain time
$t$, being the other ones decayed. In this way the causal horizon sets
up. Figure 3 and Eqs. (\ref{25}) and (\ref{26}) show that modes with larger $k$ have a longer life with reference to time t.

Since, on the other hand, longer wave-lengths correspond to lower $k$,
we see that domains with a specific spectrum of $k$-modes components
may coexist, some of them disappearing before, some other ones
persisting longer in dependence of the number in the spectrum of the
smaller or larger $k$ components, respectively. In general, the
boundaries of larger
size domains are thus expected to be less persistent 
than those of smaller size domains. This fits
with the observation \cite{BAZ}  that "critical regime has little
effect over the small scale dynamics", thus allowing the survival of
localized defects (such as vortex strings).

In conclusion, we have studied, in the framework of the GL harmonic
potential approximation, the non-equilibrium symmetry breaking phase
transition processes occurring in a finite time interval. We have
presented a possible modeling of the time-dependence of the frequency
of the order parameter
modes suitable to account for the formation of correlated domains. An interesting spectrum of possibilities for the size
and the life-time of these domains emerges, which appears to be consistent
with the conclusions of other, differently grounded analysis.

\acknowledgements

We acknowledge partial support from MURST, INFN, INFM and from the ESF
Network COSLAB.


\newpage

\begin{figure}[t]
  \caption{{\bf Case a)} Behavior in time  of $m^{2}(\beta,t)$ }
\epsfig{file=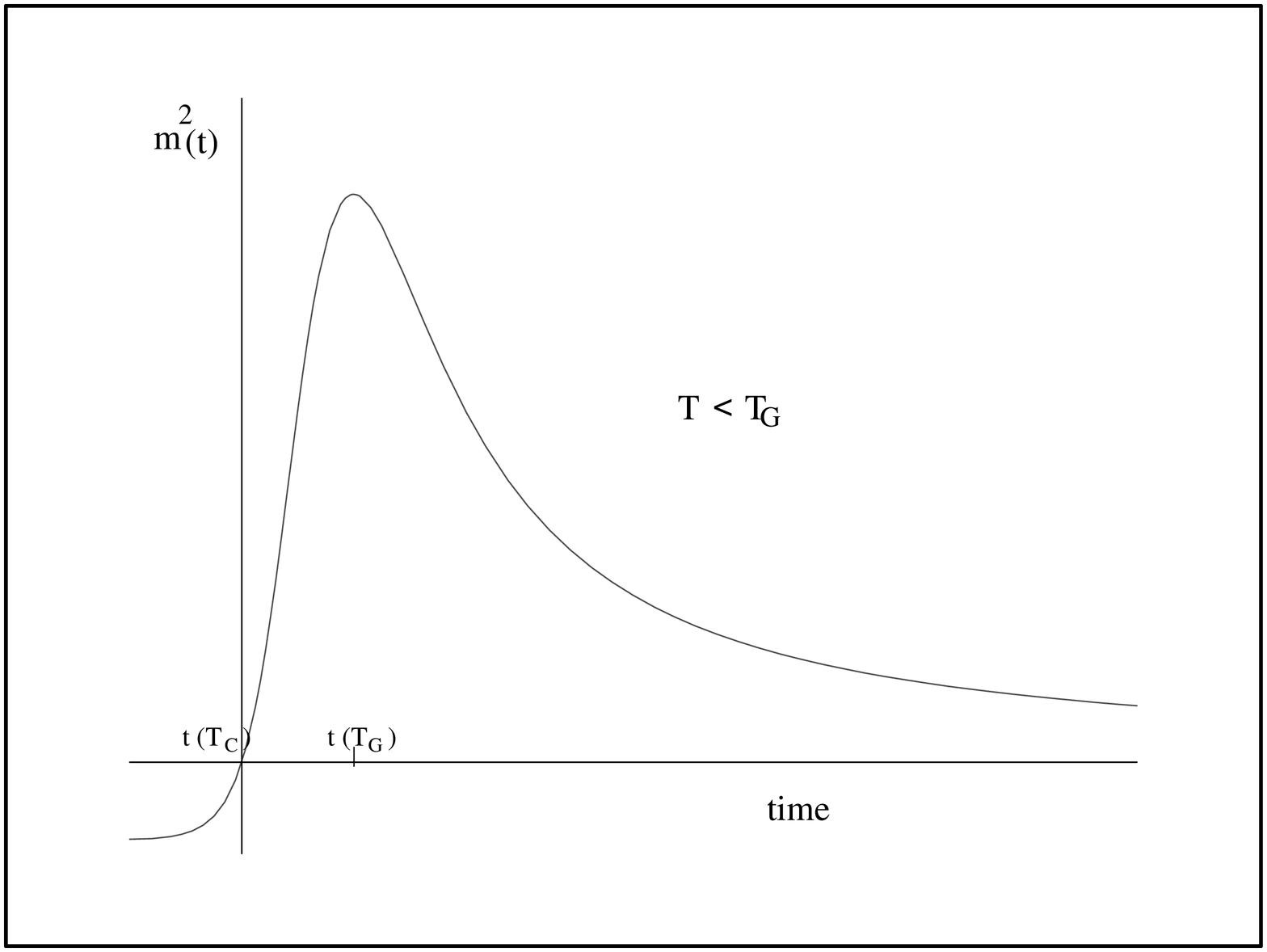,width=0.5\linewidth,height=1.0\linewidth,angle=-90}
  \end{figure}

\newpage

\begin{figure}[b]
  \caption{{\bf Case b)} Behavior in time  of $m^{2}(\beta,t)$}
\epsfig{file=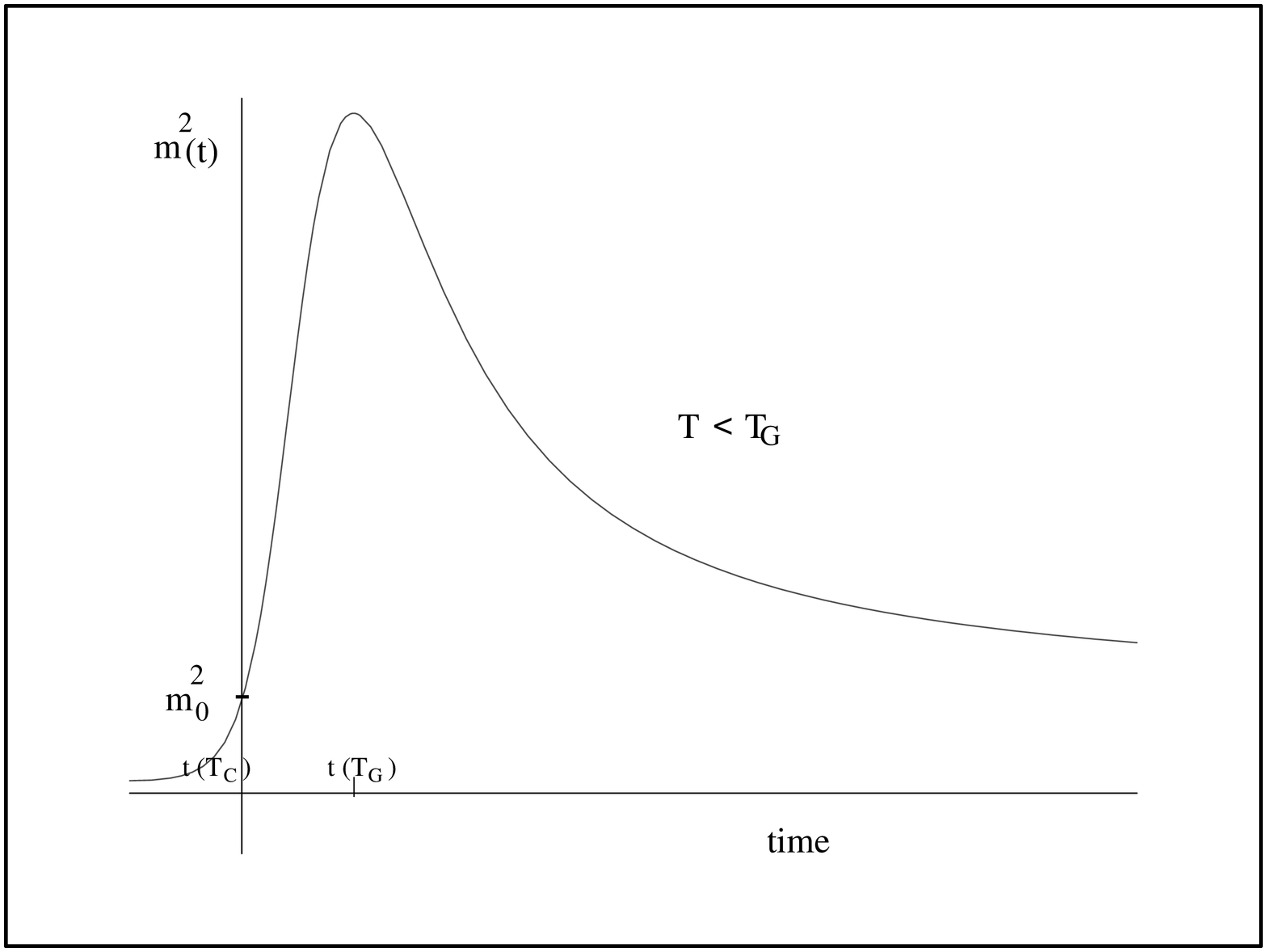,width=0.5\linewidth,height=1.0\linewidth,angle=-90}
  \end{figure}

\begin{figure}[b]
  \caption{Life-time for different values of  $k$}
\epsfig{file=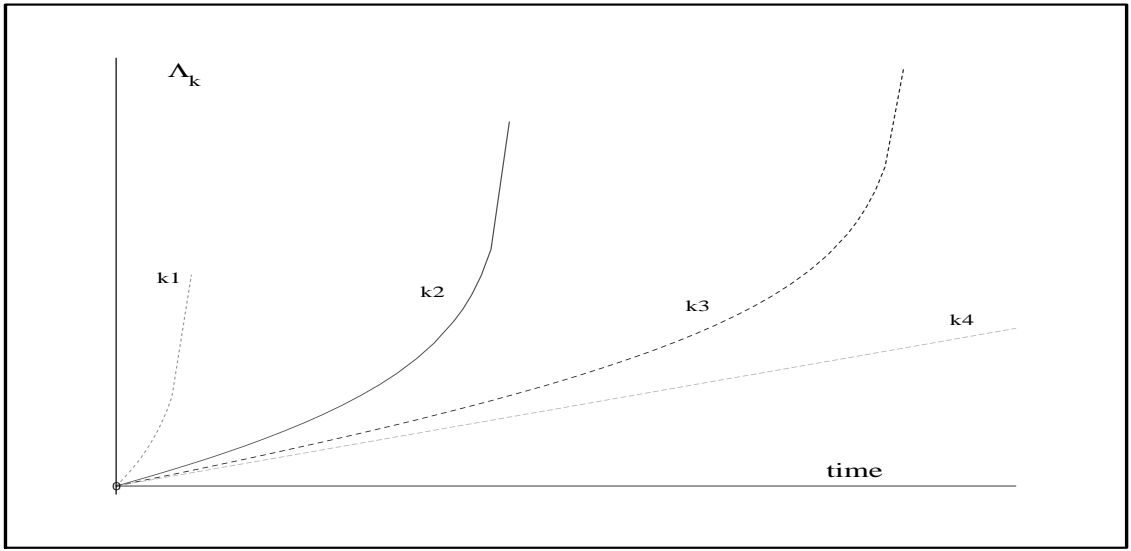,width=0.5\linewidth,height=1.0\linewidth,angle=-90}
  \end{figure}


\end{document}